\documentclass[aps,showpacs,prd,twocolumn]{revtex4}
\usepackage{amsfonts}
\usepackage{amssymb}
\usepackage{amsmath}
\usepackage{color}
\usepackage[usenames,dvipsnames]{xcolor}
\usepackage{float}
\usepackage{accents}
\usepackage{graphicx}
\usepackage{epstopdf}
\usepackage{soul}
\usepackage[colorlinks=true,pdfstartview=FitV,linkcolor=blue,citecolor=blue,urlcolor=blue,breaklinks=true]{hyperref}

\begin{document}

\title{Nontopological first-order vortices in a gauged $CP(2)$ model with a
dielectric function.}
\author{R. Casana$^{1}$, M. L. Dias$^{1}$ and E. da Hora$^{2}$.}
\affiliation{$^{1}${Departamento de F\'{\i}sica, Universidade Federal do Maranh\~{a}o,}\\
65080-805, S\~{a}o Lu\'{\i}s, Maranh\~{a}o, Brazil.\\
$^{2}$Coordenadoria Interdisciplinar de Ci\^{e}ncia e Tecnologia,\\
Universidade Federal do Maranh\~{a}o, {65080-805}, S\~{a}o Lu\'{\i}s, Maranh%
\~{a}o, Brazil{.}}

\begin{abstract}
We consider nontopological first-order solitons arising from a gauged $CP(2)$
model in the presence of the Maxwell term multiplied by a nontrivial
dielectric function. We implement the corresponding first-order scenario by
proceeding the minimization of the total energy, this way introducing the
corresponding energy lower-bound, such a construction being only possible
due to a differential constraint including the dielectric function itself
and the self-interacting potential defining the model. We saturate the
aforementioned bound by focusing our attention on those solutions fulfilling
a particular set of two coupled first-order differential equations. In the
sequel, in order to solve these equations, we choose the dielectric function
explicitly, also calculating the corresponding self-interacting potential.
We impose appropriate boundary conditions supporting nontopological
solitons, from which we verify that the energy of final structures is
proportional to the magnetic flux they engender, both quantities being not
quantized, as expected. We depict the new numerical solutions, whilst
commenting on the main properties they present.
\end{abstract}

\maketitle

\section{Introduction}

\label{Intro}

In the context of classical field theories, vortices are planar solutions
coming from highly nonlinear gauged models \cite{n5}. In general, these
solutions are calculated directly from the second-order Euler-Lagrange
equations. However, under very special circumstances, they can also be
obtained via a particular set of first-order differential equations which
minimize the energy of the overall system \cite{n4}. In this case, the
energy bound is expected to be proportional to the quantized magnetic flux
the resulting first-order vortices engender. In particular, first-order
vortices supporting quantized flux were firstly found within the canonical
Maxwell-Higgs scenario, which gives rise to topological configurations only
\cite{n1}. In the sequel, both topological and nontopological vortices were
also verified to exist in the Chern-Simons-Higgs electrodynamics \cite{cshv}.

Moreover, solitons inherent to nonstandard models were recently
investigated, for instance, the ones arising from generalizations of the
Abelian-Higgs systems \cite{gaht}, Lorentz-breaking scenarios \cite{lvs} and
gauged theories presenting nonusual dynamics whose solutions were applied to
study some interesting cosmological issues \cite{ames}.

Therefore, it is certainly important to consider whether a gauged $CP(N-1)$
theory supports well-behaved first-order vortices, especially given that the
$CP(N-1)$ model effectively mimics interesting properties of the Yang-Mills
theories defined in four-dimensions \cite{cpn-1}.

In a recent work \cite{loginov}, vortex solutions inherent to a gauged $%
CP\left( 2\right) $ model in the presence of the Maxwell term were
considered, the corresponding solutions being obtained directly from the
second-order Euler-Lagrange equations of motion. In addition, it was
suggested the existence of configurations satisfying a particular set of
first-order differential equations. In the sequel, in the contribution \cite%
{casana}, it has been developed the self-dual framework giving rise to the
aforecited solutions, whilst introducing the corresponding first-order
differential equations and the energy lower-bound explicitly. Moreover, it
has beed verified that the energy lower-bound the first-order solitons
saturate is proportional to their magnetic flux, being quantized according
the winding number rotulating such configurations, as expected. Here, it is
important to say that, due to the boundary conditions fulfilling the
finite-energy requirement, such first-order solitons present topological
properties.

In the context of noncanonical models, a rather natural issue is the study
of the gauged $CP\left( 2\right) $ theory endowed with the Maxwell term
multiplied by a dielectric function depending on the scalar field only, the
motivation regarding this nontrivial coupling coming from supersymmetric
scenarios, in which such a nonstandard kinetic term is necessary to support
a gauged model with a noncompact gauge group \cite{prd1}. Also, field models
with a dielectric function have additionally being used to study quarks and
gluons via soliton bag theories \cite{11}.

It is known that, under special circumstances, a gauged theory provided with
a nontrivial dielectric function can support both topological or
nontopological solitons. In this sense, the aim of the present manuscript is
to investigate the way such a noncanonical $CP\left( 2\right) $ model
generates nontopological solitons satisfying a particular set of first-order
differential equations.

In order to introduce our results, this manuscript is organized as follows:
in the Section II, we introduce the overall $CP(N-1)$ model and the
definitions inherent to it, from which we verify that $A^{0}=0$ holds as a
legitimate gauge choice (the theory then engendering configurations with no
electric field). In the sequel, for the sake of simplicity, we particularize
our study to the $N=3$ case, whilst focusing on those static solutions
possessing radial symmetry. We then proceed the minimization of the total
energy, from which we find the corresponding first-order equations and the
lower-bound for the total energy, such a theoretical construction being only
possible due to a differential constraint including the dielectric function
and the self-interacting potential defining the effective scenario. In the
Sec. III, we use the first-order differential equations we have found
previously to calculate genuine nontopological gauged solitons. The point to
be highlighted here is the absence of nontopological profiles for $G\left(
\left\vert \phi \right\vert \right) =1$ (the energy of the resulting
structures vanishing identically). We can contour this problem by
considering convenient nontrivial forms of the dielectric function. Also,
despite the apparent existence of two different solutions, we verify that
the effective theory provides a unique phenomenology, at least regarding the
nontopological solitons at the classical level. We present the solutions
themselves in the Section IV, pointing out the main properties they
engender. In the Section V, we end our manuscript by presenting the final
comments and perspectives regarding future contributions.

In what follows, we use $\eta ^{\mu \nu }=\left( +--\right) $ as the metric
signature and the natural units system, for convenience.

\section{The overall model \label{2}}

\label{general}

We begin by presenting the Lagrange density of the gauged $CP(N-1)$ model we
consider in this manuscript,%
\begin{equation}
\mathcal{L}=-\frac{G\left( \left\vert \phi \right\vert \right) }{4}F_{\mu
\nu }F^{\mu \nu }+\left( P_{ab}D_{\mu }\phi _{b}\right) ^{\ast }P_{ac}D^{\mu
}\phi _{c}-V\left( \left\vert \phi \right\vert \right) \text{,}  \label{1a}
\end{equation}%
where Greek indexes running over time-space coordinates, the Latin ones
representing the internal indexes of the complex $CP(N-1)$ field. Here, $%
F_{\mu \nu }=\partial _{\mu }A_{\nu }-\partial _{\nu }A_{\mu }$ is the
standard electromagnetic field strength tensor, $P_{ab}=\delta
_{ab}-h^{-1}\phi _{a}\phi _{b}^{\ast }$ being a projection operator
introduced in a convenient way. Moreover, $D_{\mu }\phi _{a}=\partial _{\mu
}\phi _{a}-igA_{\mu }Q_{ab}\phi _{b}$ stands for the covariant derivative, $%
Q_{ab}$ being a real and diagonal charge matrix. The function $G\left(
\left\vert \phi \right\vert \right) $ multiplying the Maxwell term stands
for a dielectric function to be chosen conveniently later below, the
resulting model standing for an effective action describing a system in a
medium defined by such a dielectric function. The $CP(N-1)$ field $\phi $
itself is constrained to satisfy $\phi _{a}^{\ast }\phi _{a}=h$.

The static Gauss law inherent to the model (\ref{1a}) reads%
\begin{equation}
\partial _{j}\left( G\partial ^{j}A^{0}\right) =J^{0}\text{,}  \label{2aa}
\end{equation}%
($j$ running over spatial coordinates only) with $J^{0}$, the charge
density, given by%
\begin{equation}
J^{0}=ig[(P_{ab}D^{0}\phi _{b})^{\ast }P_{ac}Q_{cd}\phi _{d}-P_{ab}D^{0}\phi
_{b}(P_{ac}Q_{cd}\phi _{d})^{\ast }],
\end{equation}%
where $D^{0}\phi _{b}=-igQ_{bc}\phi _{c}A^{0}$. It is evident that the gauge
$A^{0}=0$ satisfies (\ref{2aa}) identically. Therefore, it is possible to
infer that the resulting time-independent solutions do not present electric
field.

In what follows, we particularize our investigation to the case of the
gauged $CP(2)$ model, for a sake of simplicity. Then, we focus our attention
on those time-independent radially symmetric configurations defined by the
following ansatz:%
\begin{equation}
A_{i}=-\frac{1}{gr}\epsilon ^{ij}n^{j}A(r)\text{,}  \label{a1aa}
\end{equation}%
\begin{equation}
\left(
\begin{array}{c}
\phi _{1} \\
\phi _{2} \\
\phi _{3}%
\end{array}%
\right) =h^{\frac{1}{2}}\left(
\begin{array}{c}
e^{im_{1}\theta }\sin \left( \alpha (r)\right) \cos \left( \beta (r)\right)
\\
e^{im_{2}\theta }\sin \left( \alpha (r)\right) \sin \left( \beta (r)\right)
\\
e^{im_{3}\theta }\cos \left( \alpha (r)\right)%
\end{array}%
\right) \text{,}  \label{a2aa}
\end{equation}%
where $m_{1}$, $m_{2}$ and $m_{3}\in \mathbb{Z}$ are winding numbers, $%
\epsilon ^{ij}$ standing for the bidimensional Levi-Civita symbol (with $%
\epsilon ^{12}=+1$), $n^{j}=\left( \cos \theta ,\sin \theta \right) $ being
the position unit vector.

In such a context, regular configurations possessing no divergences are
attained via those profile functions $\alpha (r)$ and $A(r)$ fulfilling%
\begin{equation}
\alpha (r\rightarrow 0)\rightarrow 0\text{ \ \ and \ \ }A(r\rightarrow
0)\rightarrow 0\text{.}  \label{bcaa}
\end{equation}%
Moreover, given that we are interested in nontopological solitons, the
profile functions $\alpha (r)$\ and $A(r)$\ must satisfy%
\begin{equation}
\alpha (r\rightarrow \infty )\rightarrow 0\text{ \ \ and \ \ }A^{\prime
}(r\rightarrow \infty )\rightarrow 0\text{,}  \label{x12aa}
\end{equation}%
with $A(r\rightarrow \infty )\equiv A_{\infty }$ finite and arbitrary.

At this point, it is important to clarify that, regarding the charge matrix $%
Q_{ab}$ and the winding numbers $m_{1}$, $m_{2}$ and $m_{3}$, there are two
different combinations supporting first-order solutions: (i) $Q=\lambda
_{3}/2$ and $m_{1}=-m_{2}=m$, and (ii) $Q=\lambda _{8}/2$ and $m_{1}=m_{2}=m$
(both ones with $m_{3}=0$, $\lambda _{3}$ and $\lambda _{8}$ standing for
the diagonal Gell-Mann matrices: $\lambda _{3}=$diag$\left( 1,-1,0\right) $
and $\sqrt{3}\lambda _{8}=$diag$\left( 1,1,-2\right) $). However, it is
known that these two combinations are phenomenologically equivalent since
they simply mimic each other, therefore existing only one effective
scenario, as demonstrated in the Ref. \cite{loginov}. Hence, in this work,
we consider only the first choice (i.e. $m_{1}=-m_{2}=m$, $m_{3}=0$ and $%
Q=\lambda _{3}/2$), for convenience.

The second-order Euler-Lagrange equation for the additional profile function
$\beta (r) $ is%
\begin{equation}
\frac{d^{2}\beta }{dr^{2}}+\left( \frac{1}{r}+2\frac{d\alpha }{dr}\cot
\alpha \right) \frac{d\beta }{dr}=\frac{\sin ^{2}\alpha \sin \left( 4\beta
\right) }{r^{2}}\left( m-\frac{A}{2}\right) ^{2}\text{.}
\end{equation}%
We are interested in solutions with $\beta $ being a constant so such
solutions are ($k\in Z$)
\begin{equation}
\beta (r) =\beta _{1}=\frac{\pi }{4}+\frac{\pi }{2}k\text{ \ \ or \ \ }\beta
(r) =\beta _{2}=\frac{\pi }{2}k\text{,}  \label{3aa}
\end{equation}%
which apparently split our investigation in two distinct branches. However,
it is important to say that, when concerning topological first-order
solitons, these two branches engender the very same phenomenology, being
then physically equivalent. Below, we demonstrate that such equivalence also
holds regarding nontopological solitons.

We now look for first-order differential equations providing genuine
solutions of the model (\ref{1a}) by proceeding the minimization of its
total energy. In this case, given the radially symmetric ansatz (\ref{a1aa})
and (\ref{a2aa}), the effective energy reads%
\begin{eqnarray}
\frac{E}{2\pi } &=&\int \left( \frac{1}{2}GB^{2}+V\right) rdr  \notag \\
&&\hspace{-1cm}+h\int \left[ \left( \frac{d\alpha }{dr}\right) ^{2}+\frac{W}{%
r^{2}}\left( \frac{A}{2}-m\right) ^{2}\sin ^{2}\alpha \right] rdr\text{,}
\label{31aa}
\end{eqnarray}%
with $W=W\left( \alpha ,\beta \right) =1-\sin ^{2}\alpha \cos ^{2}\left(
2\beta \right) $, the function $\beta (r)$\ being necessarily one of those
presented in (\ref{3aa}), $B(r)=-A^{\prime }/gr$ standing for the magnetic
field (here, prime denotes derivative with respect to the radial coordinate $%
r$). We then write the expression (\ref{31aa}) in the form%
\begin{eqnarray}
\frac{E}{2\pi } &=&\int \!\left[ \frac{G}{2}\left( B\mp \sqrt{\frac{2V}{G}}%
\right) ^{2}\pm B\sqrt{2GV}\right] rdr  \notag \\
&&\hspace{-1cm}+h\!\!\int \!\left[ \frac{d\alpha }{dr}\mp \frac{\sqrt{W}}{r}%
\left( \frac{A}{2}-m\right) \sin \alpha \right] ^{2}rdr  \label{xee} \\
&&\hspace{-1cm}\mp \!\!\int \!\!\left[ \!\frac{d(A-2m)}{dr}\frac{\sqrt{2GV}}{%
g}+(A-2m)h\sqrt{W}\frac{d\cos \alpha }{dr}\!\right] dr\text{.}  \notag
\end{eqnarray}

In what follows, we impose the constraint%
\begin{equation}
\frac{d}{dr}\left( \sqrt{2GV}\right) =gh\sqrt{W}\frac{d}{dr}\cos \alpha
\text{,}  \label{41aa}
\end{equation}%
via which we rewrite (\ref{xee}) as%
\begin{eqnarray}
E &=&E_{bps}+\pi \int G\left( B\mp \sqrt{\frac{2V}{G}}\right) ^{2}rdr  \notag
\\
&&\hspace{-0.5cm}+2\pi h\int \left[ \frac{d\alpha }{dr}\mp \frac{\sqrt{W}}{r}%
\left( \frac{A}{2}-m\right) \sin \alpha \right] ^{2}rdr\text{,}  \label{4aa}
\end{eqnarray}%
with the energy bound $E_{bps}$\ given by%
\begin{equation}
E_{bps}=2\pi \int r\varepsilon _{bps}dr\text{,}  \label{5aax}
\end{equation}%
where%
\begin{equation}
\varepsilon _{bps}=\mp \frac{2}{gr}\frac{d}{dr}\left[ \left( \frac{A}{2}%
-m\right) \sqrt{2GV}\right] \text{,}  \label{5aa0}
\end{equation}%
the upper (lower) sign holding for negative (positive) values of $m$. In
this case, the lower-bound (\ref{5aax}) can be easily calculated, i.e.%
\begin{equation}
E_{bps}=\mp \frac{2\pi }{g}\left[ \left( A_{\infty }-2m\right) \sqrt{%
2G_{\infty }V_{\infty }}+2m\sqrt{2G_{0}V_{0}}\right] \text{,}  \label{5aa2}
\end{equation}%
where $G_{0}\equiv G(r\rightarrow 0)$, $V_{0}\equiv V(r\rightarrow 0)$, $%
G_{\infty }\equiv G(r\rightarrow \infty )$\ and $V_{\infty }\equiv
V(r\rightarrow \infty )$, with $G_{0}V_{0}$\ and $G_{\infty }V_{\infty }$\
nonnegative and finite.

Now, given the expression (\ref{4aa}), one concludes that the field
solutions saturating the energy lower-bound (\ref{5aa2}) are those ones
satisfying%
\begin{equation}
B=\pm \sqrt{\frac{2V}{G}}\text{,}  \label{bps1aa}
\end{equation}%
and%
\begin{equation}
\frac{d\alpha }{dr}=\pm \frac{\sin \alpha }{r}\left( \frac{A}{2}-m\right)
\sqrt{1-\sin ^{2}\alpha \cos ^{2}\left( 2\beta \right) }\text{,}
\label{bps2aa}
\end{equation}%
which stand for the first-order equations related to the effective radially
symmetric scenario. Here, it is important to highlight that the self-duality
supporting the first-order equations (\ref{bps1aa}) and (\ref{bps2aa}) holds
only in the presence of the constraint (\ref{41aa}), such a constraint
allowing us to rewrite the energy (\ref{31aa}) in the form (\ref{4aa}) via
the Bogomol'nyi prescription \cite{n4}.

It is also interesting to calculate the magnetic flux $\Phi _{B}$\ the
radially symmetric configurations support. It reads%
\begin{equation}
\Phi _{B}=2\pi \int rB(r)dr=-\frac{2\pi }{g}A_{\infty }\text{,}  \label{6aa}
\end{equation}%
with $A_{\infty }\equiv A(r\rightarrow \infty )$ being not necessarily
proportional to the winding number $m$, both the magnetic flux and the
energy lower-bound (\ref{5aa2}) \ being not quantized.

In the next Section, we show how the first-order expressions we have
introduced above can be used to generate well-behaved nontopological
structures possessing finite-energy, these configurations satisfying the
radially symmetric Euler-Lagrange equations, this way standing for
legitimate solutions of the corresponding model. In this manuscript, for
simplicity, we consider those cases fulfilling $G_{0}V_{0}=G_{\infty
}V_{\infty }$, the energy of the resulting configurations equaling%
\begin{equation}
E=E_{bps}=\mp \frac{2\pi }{g}A_{\infty }\sqrt{2G_{\infty }V_{\infty }}\text{,%
}  \label{5aa}
\end{equation}%
with $G_{\infty }V_{\infty }$\ being positive and finite.

%%%%%%%%%%%%%%%%%%%%%%%%

\section{Nontopological first-order scenarios}

We now go further in our investigation by using the first-order framework we
have introduced previously to obtain finite-energy nontopological solitons.
We proceed according the prescription: firstly, we pick a particular
solution for the function $\beta (r)$ coming from (\ref{3aa}), from which we
solve the differential constraint (\ref{41aa}) in order to get a concrete
relation between the dielectric function $G\left( \left\vert \phi
\right\vert \right) $ and the self-interacting potential $V\left( \left\vert
\phi \right\vert \right) $. We then choose $G\left( \left\vert \phi
\right\vert \right) $ conveniently, this way getting the potential $V\left(
\left\vert \phi \right\vert \right) $ defining that particular model, also
writing down the corresponding first-order equations (\ref{bps1aa}) and (\ref%
{bps2aa}). We particularize the expression for the radially symmetric energy
density coming from (\ref{31aa}), the functions $\alpha (r)$ and $A(r)$
obeying the asymptotic boundary conditions (\ref{x12aa}). Finally, we use
such conditions together with those in (\ref{bcaa}) in order to solve the
first-order differential equations numerically, from which we depict the
resulting profiles for $\alpha (r)$, $A(r)$, the magnetic field and energy
density they engender. We also calculate the corresponding total energy (\ref%
{5aa}) and magnetic flux (\ref{6aa}) explicitly.

It is important to discuss the absence of nontopological solitons within the
usual model, i.e. for $G\left( \left\vert \phi \right\vert \right) =1$. In
this case, the energy lower-bound (\ref{5aa}) reduces to $E=E_{bps}=\mp 2\pi
g^{-1}A_{\infty }\sqrt{2V_{\infty }}$. The point to be raised is that, in
order to fulfill the finite-energy requirement $\varepsilon (r\rightarrow
\infty )\rightarrow 0$, the self-interacting potential must satisfy $%
V_{\infty }\equiv V(r\rightarrow \infty )\rightarrow 0$, from which one also
gets $E_{bps}=0$, the corresponding solutions being energetically irrelevant.

Therefore, in this work, in order to avoid the aforementioned scenario, we
consider nontrivial expressions for the dielectric function $G\left(
\left\vert \phi \right\vert \right) $.

In the sequel, we study the cases with $\beta (r) =\beta _{1}$ and $\beta
(r) =\beta _{2}$ separately.

\subsection{The $\protect\beta (r)=\protect\beta _{1}$ case}

It was demonstrated recently \cite{casana} that such a case gives rise to
well-behaved first-order topological solitons with radial symmetry. Now, we
go a little bit further by investigating the nontopological structures $%
\beta (r)=\beta _{1}$ supports. In this sense, we choose%
\begin{equation}
\beta (r)=\beta _{1}=\frac{\pi }{4}+\frac{\pi }{2}k\text{,}  \label{v10}
\end{equation}%
via which the differential constraint (\ref{41aa}) can be reduced to%
\begin{equation}
\frac{d}{dr}\left( \sqrt{2GV}\right) =\frac{d}{dr}\left( gh\cos \alpha
\right) \text{,}  \label{v10a}
\end{equation}%
its solution defining the potential $V(\alpha )$\ in terms of the dielectric
function $G(\alpha )$, i.e.%
\begin{equation}
V(\alpha )=\frac{g^{2}h^{2}}{2G(\alpha )}\cos ^{2}\alpha \text{,}
\label{xaa}
\end{equation}%
where the integration constant was conveniently set to be null.

Here, given the expression (\ref{xaa}), one notes that $G=1$\ (standard
case, absence of the dielectric function) leads to a self-interacting
potential possessing no symmetric vacuum, therefore giving rise to
topological configurations only. In this sense, the dielectric function in (%
\ref{xaa}) must be chosen in order to engender a potential exhibiting a
symmetric vacuum, such a symmetric point supporting the existence of
nontopological solitons. We then proceed by fixing%
\begin{equation}
G(\alpha )=\frac{\left( \cos \alpha \right) ^{2-2M}}{1-\cos \alpha }\text{,}
\label{x1aa}
\end{equation}%
where $M=1$, $2$, $3$ and so on. In this case, we get the potential%
\begin{equation}
V(\alpha )=\frac{g^{2}h^{2}}{2}\left( \cos \alpha \right) ^{2M}\left( 1-\cos
\alpha \right) \text{,}  \label{x2aa}
\end{equation}%
which is positive for all values of the parameter $M$\ (the resulting energy
being itself positive, therefore justifying the way $M$\ enters the
definition (\ref{x1aa})), the general first-order equations (\ref{bps1aa})
and (\ref{bps2aa}) being reduced to%
\begin{equation}
\frac{1}{r}\frac{dA}{dr}=\pm \lambda ^{2}\left( \cos \alpha \right)
^{2M-1}\left( \cos \alpha -1\right) \text{,}  \label{x11}
\end{equation}%
\begin{equation}
\frac{d\alpha }{dr}=\pm \frac{\sin \alpha }{r}\left( \frac{A}{2}-m\right)
\text{,}  \label{x12}
\end{equation}%
respectively, the parameter $\lambda $ standing for%
\begin{equation}
\lambda =\sqrt{g^{2}h}\text{.}
\end{equation}

We summarize the scenario as follows: given the dielectric function (\ref%
{x1aa}) and the self-interacting potential (\ref{x2aa}), the gauged $%
CP\left( N-1\right) $ model (\ref{1a}) (with $N=3$) supports radially
symmetric time-independent solitons of the form (\ref{a1aa}) and (\ref{a2aa}%
) (with $\beta (r)$\ as in (\ref{v10})) satisfying the first-order equations
(\ref{x11}) and (\ref{x12}), whilst behaving according the boundary
conditions in (\ref{bcaa}) and (\ref{x12aa}). Here, it is worthwhile to
point out that, for fixed values of $M$\ and $\lambda $, the equations (\ref%
{x11}) and (\ref{x12}) support well-behaved solutions only for those values
of $m$ for which the condition $0\leq \alpha \left( r\right) <\pi /2$\ is
fulfilled.

The energy density for such nontopological solitons satisfying the
first-order differential equations is%
\begin{eqnarray}
\varepsilon (r) &=&g^{2}h^{2}\left( 1-\cos \alpha \right) \left( \cos \alpha
\right) ^{2M}  \notag \\[0.15cm]
&&+2h\frac{\sin ^{2}\alpha }{r^{2}}\left( \frac{A}{2}-m\right) ^{2}\text{,}
\end{eqnarray}%
being explicitly positive, whilst attaining the finite-energy condition $%
\varepsilon (r\rightarrow \infty )\rightarrow 0$.

It is interesting to investigate the way the profile functions $\alpha
\left( r\right) $\ and $A\left( r\right) $ behave near the boundaries. In
this sense, we linearize the first-order equations (\ref{x11}) and (\ref{x12}%
), from which we calculate the behavior of these functions near the origin,
i.e.,
\begin{eqnarray}
\alpha \left( r\right) &\approx &C_{m}\left( \lambda r\right) ^{m}-\frac{%
C_{m}^{3}}{16(m+1)^{2}}\left( \lambda r\right) ^{3m+2}+\text{...,}\quad
\label{g4} \\[0.2cm]
A\left( r\right) &\approx &\frac{C_{m}^{2}}{4\left( m+1\right) }\left(
\lambda r\right) ^{2m+2}  \notag \\[0.15cm]
&&-\frac{\left( 12M-5\right) C_{m}^{4}}{48(2m+1)}\left( \lambda r\right)
^{4m+2}+\text{...,}  \label{g5}
\end{eqnarray}%
the asymptotic behavior (i.e. for $r\rightarrow \infty $) reading%
\begin{eqnarray}
\alpha \left( r\right) &\approx &\frac{C}{\left( \lambda r\right) ^{\delta
_{m}}}+\text{...,}\quad  \label{g6} \\[0.15cm]
A\left( r\right) &\approx &2m+2\delta _{m}-\frac{C^{2}}{4\left( \delta
_{m}-1\right) \left( \lambda r\right) ^{2\delta _{m}-2}}+\text{...,}\qquad
\label{g7}
\end{eqnarray}%
where we have set $A_{\infty }\equiv A\left( r\rightarrow \infty \right)
=2m+2\delta _{m}$, $C_{m}$, $C$ and $\delta _{m}$ standing for real and
positive constants to be determined numerically by requiring the proper
behavior near the origin and asymptotically, respectively.

It is known that the solution $\beta (r)=\beta _{2}=\pi k/2$ gives rise to
first-order topological configurations possessing finite-energy. Next, we
discuss the way such a solution engenders nontopological solitons as well.

\subsection{The $\protect\beta (r)=\protect\beta _{2}$ case}

We now investigate the first-order nontopological configurations that $\beta
(r)=\beta _{2}$ supports. In this sense, we choose%
\begin{equation}
\beta (r)=\beta _{2}=\frac{\pi }{2}k\text{,}  \label{carai}
\end{equation}%
the constraint (\ref{41aa}) being rewritten in the form%
\begin{equation}
\frac{d}{dr}\left( \sqrt{2GV}\right) =\frac{d}{dr}\left( \frac{gh}{2}\cos
^{2}\alpha \right) \text{,}
\end{equation}%
\begin{figure}[tbp]
\centering\includegraphics[width=8.5cm]{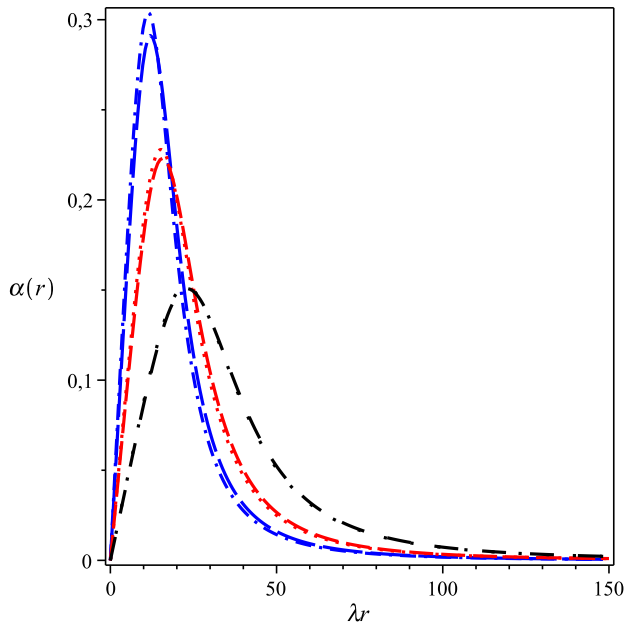}
\par
\vspace{-0.3cm}
\caption{Solutions to $\protect\alpha (r)$ for $\protect\lambda r_{0}=15$
(blue lines), $\protect\lambda r_{0}=20$ (red lines) and $\protect\lambda %
r_{0}=30$ (black lines). Here, $M=g=h=m=1$, the dashed lines standing for
the numerical solutions, the dotted lines representing the approximate ones
coming from (\protect\ref{l5}). The profiles are rings centered at the
origin.}
\end{figure}
whose solutions is%
\begin{equation}
V(\alpha )=\frac{g^{2}h^{2}}{32G(\alpha )}\cos ^{2}\left( 2\alpha \right)
\text{,}
\end{equation}%
where the integration constant was chosen to be $-gh/4$. Again, we have
found a relation between the dielectric function and the potential defining
the model.

We proceed in the very same way as before, i.e. in order to have a potential
with a symmetric vacuum (therefore supporting nontopological profiles, see
the discussion just before the Eq. (\ref{x1aa})), we choose the dielectric
function as%
\begin{equation}
G(\alpha )=\frac{\left( \cos \left( 2\alpha \right) \right) ^{2-2M}}{1-\cos
\left( 2\alpha \right) }\text{,}  \label{yaa}
\end{equation}%
then getting (as we demonstrate below, the factor $2\alpha $\ was introduced
in (\ref{yaa}) in order to make the two a priori different scenarios
phenomenologically equivalent via a suitable redefinition)%
\begin{equation}
V(\alpha )=\frac{g^{2}h^{2}}{32}\left( \cos \left( 2\alpha \right) \right)
^{2M}\left( 1-\cos \left( 2\alpha \right) \right) \text{,}  \label{ybb}
\end{equation}%
the potential itself and its energy density being positive for all $M$, the
corresponding first-order equations (\ref{bps1aa}) and (\ref{bps2aa})
thereby reading%
\begin{equation}
\frac{1}{r}\frac{dA}{dr}=\pm \frac{g^{2}h}{4}\left( \cos \left( 2\alpha
\right) \right) ^{2M-1}\left( \cos \left( 2\alpha \right) -1\right) \text{,}
\label{y11}
\end{equation}%
\begin{equation}
\frac{d\alpha }{dr}=\pm \frac{\sin \left( 2\alpha \right) }{2r}\left( \frac{A%
}{2}-m\right) \text{,}  \label{y12}
\end{equation}%
respectively. In order to obtain nontopological structures, the above
first-order equations must be solved according the boundary conditions (\ref%
{bcaa}) and (\ref{x12aa}), the energy density of the resulting solutions
being%
\begin{eqnarray}
\varepsilon (r) &=&\ \frac{g^{2}h^{2}}{16}\left( \cos \left( 2\alpha \right)
\right) ^{2M}\left( 1-\cos \left( 2\alpha \right) \right)   \notag \\[0.15cm]
&&+\frac{h}{2}\frac{\sin ^{2}\left( 2\alpha \right) }{r^{2}}\left( \frac{A}{2%
}-m\right) ^{2}\text{.}
\end{eqnarray}%
\begin{figure}[tbp]
\centering\includegraphics[width=8.5cm]{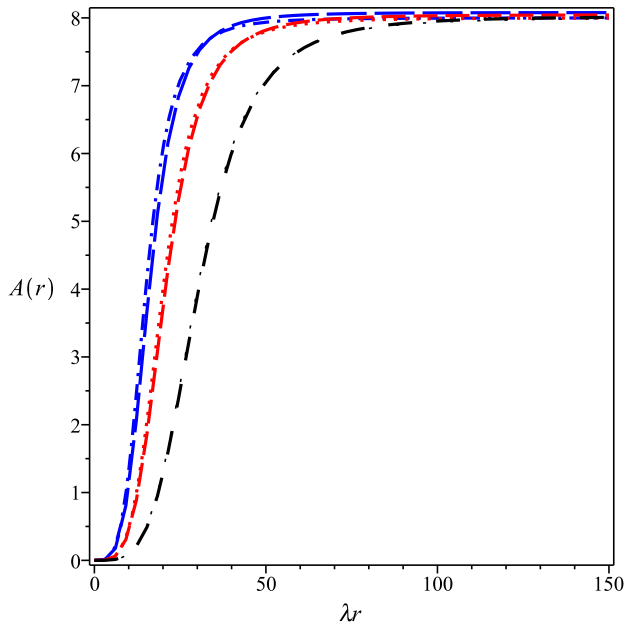}
\par
\vspace{-0.3cm}
\caption{Solutions to $A(r)$. Conventions as in the Fig. 1, the approximate
results being given by (\protect\ref{l6}). The numerical solutions approach
the approximate value $A_{m}(r\rightarrow \infty )=4\left( m+1\right) $ in a
monotonic way whenever $\protect\lambda r_{0}$ increases.}
\end{figure}

It is known that, regarding the first-order topological configurations,
those two a priori different scenarios the solutions for $\beta (r)$ in (\ref%
{3aa}) engender simply mimic each other, therefore existing only one
effective case. Here, it is important to highlight that such correspondence
also holds regarding the first-order nontopological structures we have
introduced above, i.e., both the dielectric function in (\ref{x1aa}) and the
self-interacting potential in (\ref{x2aa}) can be rewritten as those ones in
(\ref{yaa}) and (\ref{ybb}), respectively, whether we implement the
redefinitions $\alpha \rightarrow 2\alpha $ and $h\rightarrow h/4$, the
corresponding first-order equations (\ref{x11}) and (\ref{x12}) then
reducing to (\ref{y11}) and (\ref{y12}), this way existing only one
effective scenario, at least regarding the nontopological radially symmetric
first-order solitons in the presence of a nontrivial dielectric function.

Therefore, from now on, we investigate only the case defined by $\beta
(r)=\beta _{1}$, the resulting first-order equations being (\ref{x11}) and (%
\ref{x12}).

\section{The solutions}

It is instructive to point out that the first-order equations (\ref{x11})
and (\ref{x12}) support approximate analytical solutions describing the
corresponding nontopological configurations. In what follows, we investigate
these solutions by choosing $m>0$ only (i.e. the lower signs in the
first-order expressions), for simplicity. We also suppose that $\alpha
(r)\ll 1$ for all values of $\lambda r$. In this sense, the first-order
equations (\ref{x11}) and (\ref{x12}) reduce, respectively, to%
\begin{equation}
\frac{1}{r}\frac{dA}{dr}=\mp \frac{\lambda ^{2}}{2}\alpha ^{2}\text{,}
\label{x11a}
\end{equation}%
\begin{equation}
\frac{d\alpha }{dr}=\pm \frac{\alpha }{r}\left( \frac{A}{2}-m\right) \text{,}
\label{x12a}
\end{equation}%
which can be combined to each other into the Liouville's equation for the
profile function $\alpha (r)$, i.e.%
\begin{equation}
\frac{d^{2}\left( \ln \alpha ^{2}\right) }{dr^{2}}+\frac{1}{r}\frac{d\left(
\ln \alpha ^{2}\right) }{dr}+\frac{\lambda ^{2}}{2}\alpha ^{2}=0\text{,}
\end{equation}%
whose general solution is \cite{jackiw90}%
\begin{equation}
\alpha (r)=\frac{4C}{{\lambda }r_{0}}\frac{\left( \frac{r}{r_{0}}\right)
^{C-1}}{1+\left( \frac{r}{r_{0}}\right) ^{2C}}\text{,}  \label{xl4}
\end{equation}%
with $C$\ and $r_{0}$\ $\ $standing for integration constants. Here, it is
interesting to note that this solution satisfies the conditions $\alpha
(r\rightarrow 0)\rightarrow 0$ and $\alpha (r\rightarrow \infty )\rightarrow
0$\ for $C>1$ only.
\begin{figure}[tbp]
\centering\includegraphics[width=8.5cm]{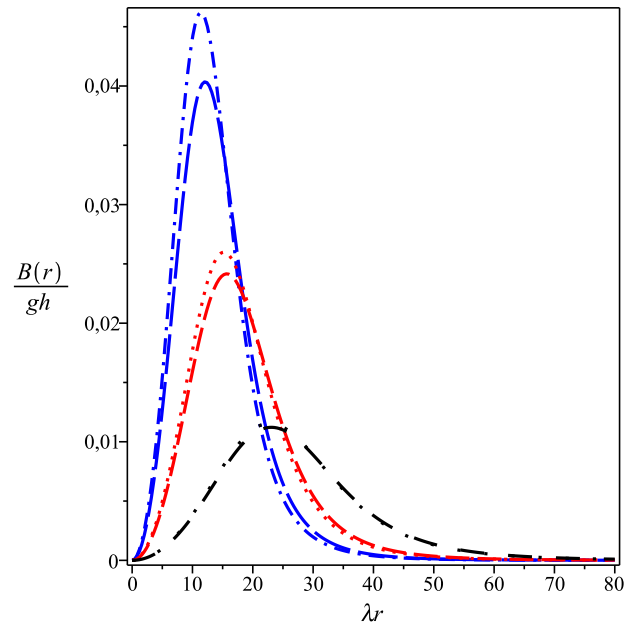}
\par
\vspace{-0.3cm}
\caption{Solutions to the magnetic field $B(r)$, in units of $gh$.
Conventions as in the Fig. 1, the approximate solutions coming from (\protect
\ref{mfaa}). Here, the corresponding amplitudes (radii) decrease (increase)
as $\protect\lambda r_{0}$ itself increases.}
\end{figure}

In addition, given the solution (\ref{xl4}), the first-order equation (\ref%
{x12a}) provides
\begin{figure}[tbp]
\centering\includegraphics[width=8.5cm]{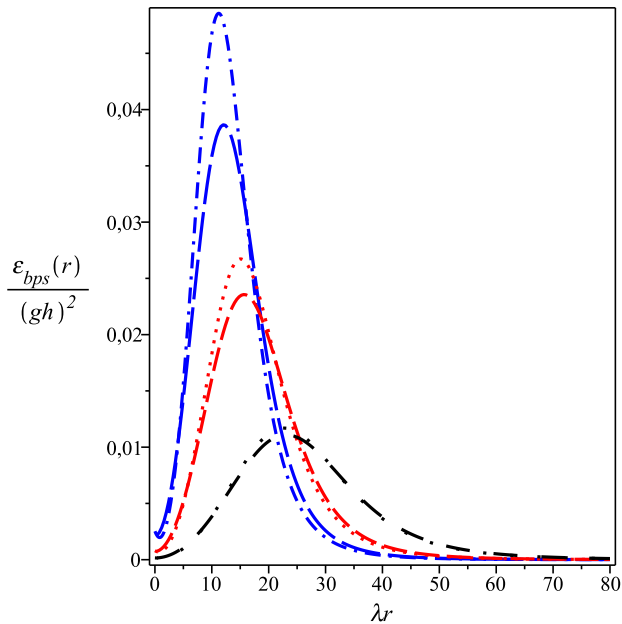}
\par
\vspace{-0.3cm}
\caption{Solutions to the energy density $\protect\varepsilon _{bps}(r)$, in
units of $\left( gh\right) ^{2}$. Conventions as in the Fig. 1, the
approximate profiles standing for (\protect\ref{edaa}). The nontopological
structures are well-localized in space.}
\end{figure}
\begin{equation}
A(r)=2\left( m+1\right) -2C+\frac{4C\left( \frac{r}{r_{0}}\right) ^{2C}}{%
1+\left( \frac{r}{r_{0}}\right) ^{2C}}\text{,}  \label{xl5}
\end{equation}%
which fulfills $A(r\rightarrow 0)\rightarrow 0$\ only for $C=m+1$.
Therefore, the approximate solutions (\ref{xl4}) and (\ref{xl5}) can be
rewritten, respectively, as%
\begin{equation}
\alpha _{m}(r)=\frac{4\left( m+1\right) }{{\lambda }r_{0}}\frac{\left( \frac{%
r}{r_{0}}\right) ^{m}}{1+\left( \frac{r}{r_{0}}\right) ^{2m+2}}\text{,}
\label{l5}
\end{equation}%
\begin{equation}
A_{m}(r)=4\left( m+1\right) \frac{\left( \frac{r}{r_{0}}\right) ^{2m+2}}{%
1+\left( \frac{r}{r_{0}}\right) ^{2m+2}}\text{,}  \label{l6}
\end{equation}%
from which one also gets approximate expressions for the magnetic field%
\begin{equation}
B_{m}(r)=-\frac{gh}{2}\alpha _{m}^{2}\text{,}  \label{mfaa}
\end{equation}%
and the energy density%
\begin{equation}
\varepsilon _{bps,m}(r)=\frac{g^{2}h}{2}\alpha _{m}^{2}+2h\frac{\alpha
_{m}^{2}}{r^{2}}\left( \frac{A_{m}}{2}-m\right) ^{2}\text{,}  \label{edaa}
\end{equation}%
with $\alpha _{m}(r)$ and $A_{m}(r)$ being given by\ (\ref{l5}) and (\ref{l6}%
), respectively, the approximate value for $A_{\infty }\equiv A(r\rightarrow
\infty )$ being calculated exactly, i.e.%
\begin{equation}
A_{\infty ,m}\equiv A_{m}(r\rightarrow \infty )=4\left( m+1\right) \text{.}
\label{l7}
\end{equation}%
\begin{figure}[tbp]
\centering\includegraphics[width=8.5cm]{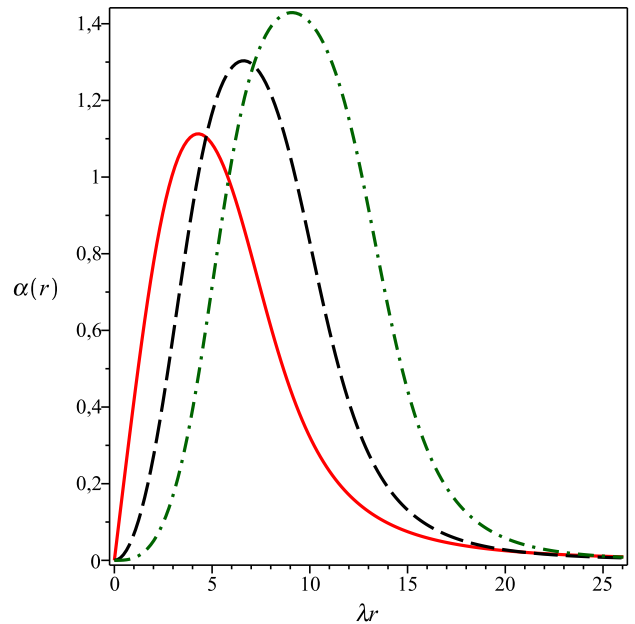}
\par
\vspace{-0.3cm}
\caption{Solutions to $\protect\alpha (r)$ for $m=1$ (solid red line), $m=2$
(dashed black line) and $m=3$ (dash-dotted green line). Now, $M=g=h=\protect%
\lambda r_{0}=1$, the profiles being again rings centered at the origin.}
\end{figure}

The function (\ref{l5}) has its maximum value given by%
\begin{equation}
\alpha _{m}(r_{\max })=\frac{2\left( m+2\right) }{{\lambda }r_{0}}\left(
\frac{m}{m+2}\right) ^{\frac{m}{2(m+1)}}\text{,}
\end{equation}%
where%
\begin{equation}
r_{\max }=r_{0}\left( \frac{m}{m+2}\right) ^{\frac{1}{2(m+1)}}\text{,}
\end{equation}%
which approximates $r_{0}$\ for large values of $m$, our previous assumption
$\alpha (r)\ll 1$\ holding whether the additional condition%
\begin{equation}
{\lambda }r_{0}\gg 2\left( m+2\right) \left( \frac{m}{m+2}\right) ^{\frac{m}{%
2(m+1)}}
\end{equation}%
is fulfilled. Therefore, for fixed values of $g$, $h$ and $r_{0}$, only some
values of the integer winding number $m$ are allowed.

We highlight that, given the dielectric function (\ref{x1aa}), the
self-interacting potential (\ref{x2aa}) and the boundary value (\ref{l7}),
the resulting energy bound (\ref{5aa}) can be calculated explicitly, being
then equal to (remenber that $m>0$)%
\begin{equation}
E_{bps}=8\pi \left( m+1\right) h\text{,}
\end{equation}%
where we have used $G_{0}V_{0}=G_{\infty }V_{\infty }=g^{2}h^{2}/2$ (this
way verifying our previous conjecture), the corresponding magnetic flux (\ref%
{6aa}) reading
\begin{equation}
\Phi _{B}=-\frac{8\pi }{g}\left( m+1\right) \text{,}
\end{equation}%
from which we get that $E_{bps}=-gh\Phi _{B}$, the energy of the first-order
solitons being then proportional to their magnetic flux, as expected.
\begin{figure}[tbp]
\centering\includegraphics[width=8.5cm]{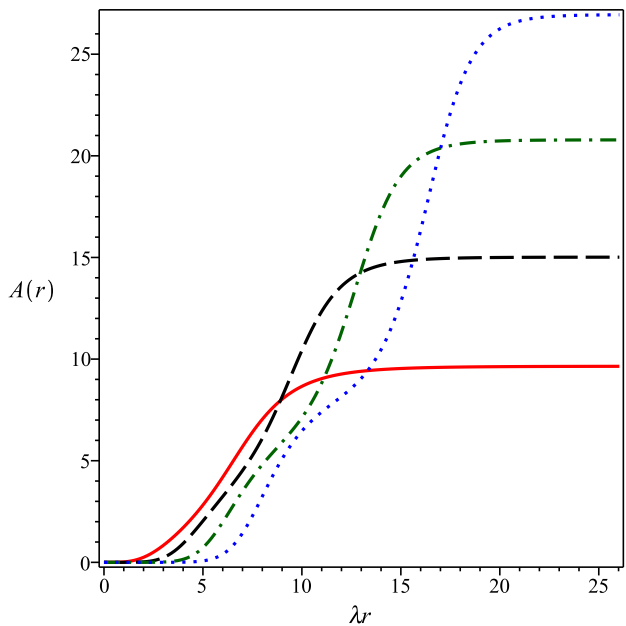}
\par
\vspace{-0.3cm}
\caption{Solutions to $A(r)$. Conventions as in the Fig. 5, the additional
dotted blue line standing for $m=4$, the corresponding solution highlighting
the existence of an internal structure.}
\end{figure}

In what follows, we proceed the numerical study of the first-order equations
(\ref{x11}) and (\ref{x12}) by means of the finite-difference scheme, whilst
using the boundary conditions (\ref{bcaa}) and (\ref{x12aa}). We adopt $m>0$
(lower signs) and $M=g=h=1$, for simplicity. In this sense, we plot the
numerical solutions to the profile functions $\alpha (r)$ and $A(r)$, the
magnetic field $B(r)$ and the energy density $\varepsilon _{bps}(r)$. In
these figures, we have used the dimensionless variable $\lambda r$ in order
to depict the numerical profiles, the only free parameter remaining being $%
\lambda r_{0}$.

We begin our analysis by depicting the numerical profiles corresponding to
the approximate expressions in (\ref{l5}) and (\ref{l6}). We choose $m=1$,
whilst varying $\lambda r_{0}$, the analytical solutions approximating the
numerical ones for large values of such a parameter. Here, the dashed lines
stand for the numerical solutions, the dotted lines representing the
approximate ones (see Figs. 1-4).

In the Figure 1, we show the solutions to the profile function $\alpha (r)$
for $\lambda r_{0}=15$ (blue lines), $\lambda r_{0}=20$ (red lines) and $%
\lambda r_{0}=30$ (black lines). We see that the resulting profiles are
rings centered at the origin, their amplitudes (radii) decreasing
(increasing) as $\lambda r_{0}$ itself increases, the numerical results
fulfilling our previous assumption (i.e. that $\alpha (r)\ll 1$ for all $%
\lambda r$), the approximate solutions fitting relatively well.

The solutions to the field $A(r)$ are those shown in the Figure 2, from
which we see that these profiles approach the approximate boundary condition
$A_{m}(r\rightarrow \infty )=4\left( m+1\right) $ in a monotonic way
whenever $\lambda r_{0}$ increases. In particular, one gets the numerical
values $A_{1}(r\rightarrow \infty )\approx 8.07855$ for $\lambda r_{0}=15$, $%
A_{1}(r\rightarrow \infty )\approx 8.04565$ for $\lambda r_{0}=20$ and $%
A_{1}(r\rightarrow \infty )\approx 8.02063$ for $\lambda r_{0}=30$.
\begin{figure}[tbp]
\centering\includegraphics[width=8.5cm]{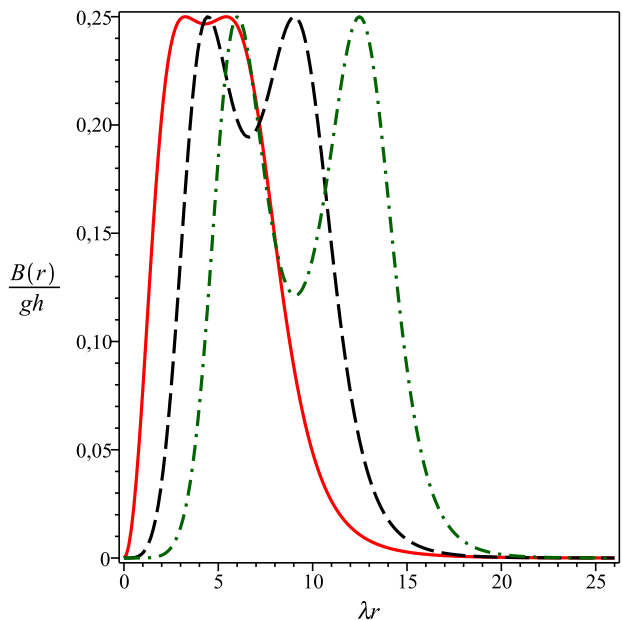}
\par
\vspace{-0.3cm}
\caption{Solutions to the magnetic field $B(r)$, in units of $gh$.
Conventions as in the Fig. 5. The new profiles are double rings centered at $%
r=0$.}
\end{figure}

In the Figure 3, we depict the numerical solutions to the magnetic field $%
B(r)$, in units of $gh$, these profiles behaving as those for $\alpha (r)$,
their amplitudes (radii) decreasing (increasing) as $\lambda r_{0}$
increases. The new solutions also vanish in the asymptotic limit $%
r\rightarrow \infty $, this way fulfilling the finite-energy requirement $%
\varepsilon (r\rightarrow \infty )\rightarrow 0$.

The solutions to the energy density $\varepsilon _{bps}(r)$ appear in the
Figure 4, in units of $\left( gh\right) ^{2}$, showing that the
corresponding nontopological structures are localized in space. Here, we
highlight the manner $\varepsilon _{bps}(r=0)$ depends on $\lambda r_{0}$,
that value increasing from 0 (zero) as this free parameter decreases.

It is also interesting to point out the existence of another class of
numerical solutions that can not be predicted by an approximate analytical
treatment, the condition $\alpha (r)\ll 1$ for all $\lambda r$ being not
satisfied anymore. These new profiles are calculated for finite, but no
large, values of the free parameter $\lambda r_{0}$, this way differing from
the solutions presented above (see Figs. 5-8).

In order to introduce the aforecited profiles, we again solve (\ref{x11})
and (\ref{x12}) numerically, for $m>0$ and $M=g=h=1$, via the conditions (%
\ref{bcaa}) and (\ref{x12aa}). However, now we choose $\lambda r_{0}=1$,
whilst varying the winding number: $m=1$ (solid red line), $m=2$ (dashed
black line) and $m=3$ (dash-dotted green line), plotting the corresponding
solutions with respect to the dimensionless variable $\lambda r$.

In this sense, the solutions to $\alpha (r)$ are those depicted in the Fig.
5, from which we see that these profiles behave in a similar manner as
before, being rings centered at the origin, both amplitudes and radii
increasing as the vorticity increases.

The Figure 6 shows the results to the function $A(r)$, the additional dotted
blue line standing for $m=4$. Here, it is worthwhile to note the appearance
of an interesting internal structure for intermediary values of the
dimensionless variable $\lambda r$. It is also interesting to highlight that
the new solutions do not fulfill the previous condition (\ref{l7}), the new
values being $A_{1}(r\rightarrow \infty )\approx 9.64900$, $%
A_{2}(r\rightarrow \infty )\approx 15.01548$, $A_{3}(r\rightarrow \infty
)\approx 20.78517$ and $A_{5}(r\rightarrow \infty )\approx 26.98683$.

The numerical profiles to the magnetic field $B(r)$ are plotted in the Fig.
7, in units of $gh$, from which we see that these solutions are drastically
different from the ones appearing in the Fig. 3, the new configurations
being double rings centered at $r=0$, the magnetic field vanishing in the
asymptotic limit.

Finally, the Fig. 8 shows the solutions to $\varepsilon _{bps}(r)$, again in
units of $\left( gh\right) ^{2}$, these profiles also standing for double
rings centered at the origin. However, in this case, the amplitude of the
inner ring is always greater than that of the outer one, $\varepsilon
_{bps}(r=0)$ vanishing for $m\neq 1$.

%%%%%%%%%%%%%%%%%%%%%%%%

\section{Final comments and perspectives}

In this manuscript, we have studied the $CP(2)$ model endowed by the Maxwell
term in the presence of an \textit{a priori} arbitrary dielectric function $%
G(|\phi |)$, from which we have attained nontopological first-order vortices
possessing finite-energy and nonquantized magnetic flux.

We have presented the particular model and the definitions inherent to it,
from which we have verified that $A^{0}=0$ satisfies the static Gauss law
identically, the resulting time-independent solutions supporting no electric
field. We have particularized our investigation to case $N=3$, for
simplicity. Then, we have focused our attention on those radially symmetric
configurations described by the \textit{Ansatz} defined in (\ref{a1aa}) and (%
\ref{a2aa}) whilst satisfying the boundary conditions (\ref{bcaa}) and (\ref%
{x12aa}). We have introduced convenient choices regarding the charges and
winding numbers inherent to the aforementioned \textit{Ansatz}. In the
sequel, we have calculated the solutions (\ref{3aa}) for the additional
profile function $\beta (r)$.
\begin{figure}[tbp]
\centering\includegraphics[width=8.5cm]{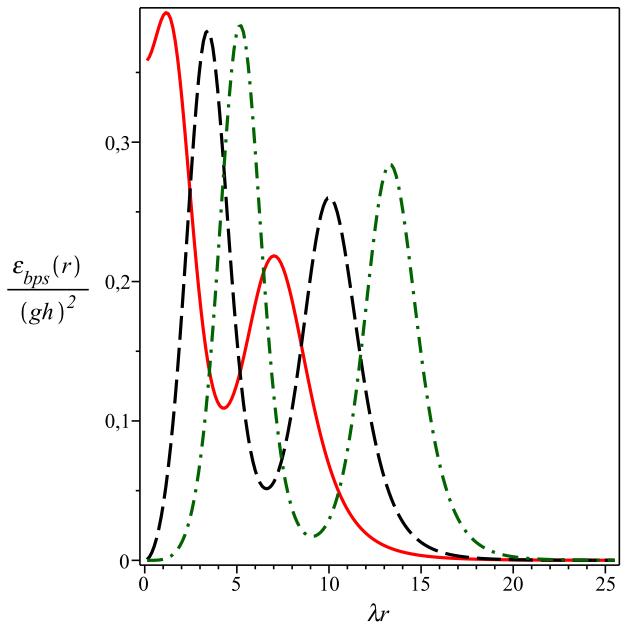}
\par
\vspace{-0.3cm}
\caption{Solutions to the energy density $\protect\varepsilon _{bps}(r)$, in
units of $\left( gh\right) ^{2}$. Conventions as in the Fig. 5. Here, $%
\protect\varepsilon _{bps}(r=0)$ vanishes for $m\neq 1$.}
\end{figure}

We have rewritten the expression for the radially symmetric energy (\ref%
{31aa}) as that in (\ref{4aa}), whilst defining the general first-order
equations (\ref{bps1aa}) and (\ref{bps2aa}) satisfied by the fields $\alpha
(r)$ and $A(r)$, additionally getting the corresponding lower-bound for the
total energy (see (\ref{4aa}) itself and (\ref{5aa})). The point to be
highlighted here is that such construction was only possible due to the
differential constraint (\ref{41aa}) including $G(|\phi |)$ and the
self-interacting potential $V(|\phi |)$. We have also calculated a general
result for the magnetic flux $\Phi _{B}$ in (\ref{6aa}).

We have discussed the absence of nontopological solitons for $G(|\phi |)=1$,
the energy of these structures vanishing. In order to avoid such scenario,
we have considered only nontrivial forms for the dielectric function.

We have studied the case defined by $\beta (r)=\beta _{1}$ (\ref{v10})
firstly, from which we have written the constraint (\ref{41aa}) as in (\ref%
{v10a}), whose solution is (\ref{xaa}). Then, we have chosen the dielectric
function as showed in (\ref{x1aa}), from which we have obtained the
potential (\ref{x2aa}) and the first-order equations (\ref{x11}) and (\ref%
{x12}). We have also linearized these equations in order to define the way
the nontopological solutions behave near the boundaries.

We have implemented the same prescription also for $\beta (r)=\beta _{2}$ (%
\ref{carai}), from which we have calculated the corresponding first-order
expressions. We have noticed that the two scenarios defined by the different
solutions for $\beta (r)$ can be verified to mimic each other via the
redefinitions $\alpha \rightarrow 2\alpha $, $\lambda \rightarrow \lambda /4$
and $h\rightarrow h/4$, therefore existing only one effective case. In this
sense, we have focused our attention on the case $\beta (r)=\beta _{1}$ only.

We have supposed that $\alpha (r)\ll 1$ for all $\lambda r$ (with $\lambda
^{2}=g^{2}h$), from which we have combined the first-order equations (\ref%
{x11}) and (\ref{x12}), therefore verifying that the function $\alpha (r)$
satisfy the Liouville equation whose analytical solution has provided the
profiles (\ref{l5}) and (\ref{l6}), also calculating the boundary value $%
A(r\rightarrow \infty )\rightarrow 4(m+1)$. In addition, we have verified
explicitly that the energy bound is proportional to the magnetic flux
inherent to the resulting solitons, both quantities being not necessarily
quantized, as expected.

We have depicted the numerical results we have found to $\alpha (r)$, $A(r)$%
, the magnetic field $B(r)$ and the energy density $\varepsilon _{bps}(r)$,
for different values of the vorticity $m$ and the parameter $\lambda r_{0}$,
from which we have pointed out the existence of two different classes of
solutions: the ones coming from large values of $\lambda r_{0}$, being
reasonably well described by the analytic expressions in (\ref{l5}) and (\ref%
{l6}), and those solutions related to small values of $\lambda r_{0}$, do
not possessing an approximate counterpart.

Here, it is important to highlight that the results we have introduced in
this work hold only for those time-independent solitons described by the
\textit{Ansatz} in (\ref{a1aa}) and (\ref{a2aa}). In this sense, it is not
possible to say that the general model (\ref{1a}) supports regular
first-order solutions outside the radially symmetric scenario, such question
lying beyond the scope of the present investigation.

Moreover, ideas regarding future works include the study of the $CP(2)$
model in the presence of the Chern-Simons action (instead of the Maxwell's
one) and the first-order configurations the resulting theory possibly
supports. This issue is now under consideration, and we hope relevant
results for an incoming contribution.

\begin{acknowledgments}
The authors thank CAPES, CNPq and FAPEMA (Brazilian agencies) for partial
financial support.
\end{acknowledgments}

\end{document}